\documentclass[10pt,a4paper]{article}
\usepackage[utf8]{inputenc}
\usepackage[T1]{fontenc}
\usepackage{amsmath}
\usepackage{amsfonts}
\usepackage{amssymb}
\usepackage{graphicx}
\usepackage[left=2.0cm, right=2.00cm, top=2.00cm, bottom=2.00cm]{geometry}	
\usepackage[hidelinks]{hyperref}
\usepackage{xcolor}

\usepackage[square,numbers]{natbib}
\usepackage{filecontents}

\usepackage{pgfplots}
\pgfplotsset{compat=newest}
\usetikzlibrary{shapes.geometric,arrows,positioning}

\author{Michael Fink\\ \small Technical University of Munich, 80333 Munich, Germany\\ \small \texttt{michael.fink@tum.de}}
\title{Implementation of Linear Model Predictive Control - Tutorial  }
\newcommand{\bs}[1]{{\ensuremath{\boldsymbol{#1}}}}

\begin{document}
\maketitle

\begin{abstract}
	This tutorial shows an overview of Model Predictive Control with a linear discrete-time system and constrained states and inputs.
	The focus is on the implementation of the method under consideration of stability and recursive feasibility. 
	The MATLAB code for the examples and plots is available online.
\end{abstract}
\tableofcontents

\newpage

\section{Introduction}
This tutorial shows a brief overview of linear Model Predictive Control (MPC) \cite{RM09}. MPC is a control method which iteratively applies optimal control. At each time instance $k$ where MPC is applied an optimal control problem is solved. Therefore, all predicted states and inputs within a prediction horizon $N$ are optimized to find an optimal input sequence. Then, only the first input of the optimal input sequence is applied to the system. This procedure is repeated at each time instance with the current state of the system as the initial state for the predictions. 
An advantage of MPC is, that it can consider constraints on the states and on the input because of the iterative solution of the optimal control problem.  
In the following, an overview of MPC is shown. Then, a reformulation is presented to formulate a quadratic program of the MPC optimization to obtain a fast computation in MATLAB.
Finally, an example of the method is shown, for which the MATLAB code is available at \url{https://github.com/placebovitamin/MPC-Tutorial}.

\section{Linear Model Predictive Control}
In this section, the fundamentals of linear MPC are shown. 

\subsection{System}
The description of the linear discrete-time system model is
	\begin{align}\label{eq:sys}
	\bs{x}_{k+1}&=\bs{A}\bs{x}_{k}+\bs{B}\bs{u}_{k}
	\end{align}	
with time step $k$, states $\bs{x}_{k}\in\mathbb{R}^n$, control input $\bs{u}_{k}\in\mathbb{R}^m$,  and matrices $\bs{A}\in\mathbb{R}^{n\times n}$, $\bs{B}\in\mathbb{R}^{n\times m}$. For the states and inputs  the following constraints hold
\begin{align}
	\bs{x}_{k} \in \mathcal{X}, \bs{u}_{k} \in \mathcal{U} \quad \forall k \in \mathbb{N}_0 ,
\end{align}
where $\mathbb{N}_0$ denotes the set of all non-negative integers.  The state set $\mathcal{X}$ is closed and the input set $\mathcal{U}$ is compact. Both sets are convex and contain the origin.

\subsection{Optimal Control Problem}
In the optimal control problem, the states are predicted based on the model \eqref{eq:sys}. In the following, the prediction  of the states at the $i$-th time steps after the current time $k$ are denoted as $\bs{\hat{x}}_{k+i|k}$. The predictions are based on the current measured state of the system \eqref{eq:sys}, which is the $k$-th state $\bs{x}_k$, i.e.,
\begin{align}\label{eq:measure}
	\bs{\hat{x}}_{k|k} = \bs{x}_{k}.
\end{align}
For the prediction of the states with the model \eqref{eq:sys} the inputs are needed. These inputs are the decision variables for the optimal control problem and therefore chosen such that the state prediction and the chosen inputs lead to an optimal solution according to a cost function. 
Based on the current state $\bs{x}_k$, the input $i$ time steps in the future is denoted as $\bs{\hat{u}}_{k+i|k}$.

\subsubsection{Regulation Problem}\label{ssec:reguation}
The goal in the regulation problem is that the inputs are determined such that the states are steered to the origin. Therefore, an objective function is utilized that considers the predictions in the next $N$ steps. Thus, $N$ is the prediction horizon of the optimization problem. 
Typically, for the states and inputs a quadratic cost function $J(\bs{\hat{x}}_{k|k}, \bs{U}_k)$ is used because an optimization with a quadratic norm is easier to compute as for example a $l_1$-norm. 
The weight matrix for states is denoted as $\bs{Q}\in\mathbb{R}^{n\times n}$ and it is positive semi definite, i.e., $\bs{Q}\succeq 0$. The weight matrix for the inputs is $\bs{R}\in\mathbb{R}^{m\times m}$ and it is positive definite, i.e.,  $\bs{R}\succ0$.
Both matrices are user-defined and typically chosen as  diagonal matrices. 
A high value in $\bs{Q}$ leads to a faster convergence of the corresponding state, whereas a high value of $\bs{R}$ reduces the amplitude of the input.  
Therefore, the finite time optimal control problem is 
\begin{subequations}\label{eq:opt}
	\begin{alignat}{2}
	\bs{U}_k^* = & \arg \min_{\bs{U}_k}  J( \bs{\hat{x}}_{k|k}, \bs{U}_k)\\ 
	=&\arg \min_{\bs{U}_k}\sum_{i=0}^{N-1} \left[\bs{\hat{x}}_{k+i|k}^\top \bs{Q} \bs{\hat{x}}_{k+i|k}+ \bs{\hat{u}}_{k+i|k}^\top \bs{R} \bs{\hat{u}}_{k+i|k} \right ] +&& \bs{\hat{x}}_{k+N|k} ^\top \bs{Q}_\text{f} \bs{\hat{x}}_{k+N|k}  \\
	\text{s.t.}\quad  &  \bs{\hat{x}}_{k+i+1|k}=\bs{A}\bs{\hat{x}}_{k+i|k}+\bs{B}\bs{\hat{u}}_{k+i|k}&& i \in \mathbb{N}_{0:N-1}  \\
	& \bs{\hat{u}}_{k+i|k} \in \mathcal{U} && i \in \mathbb{N}_{0:N-1} \\
	& \bs{\hat{x}}_{k+i|k} \in \mathcal{X} && i \in \mathbb{N}_{0:N-1}  \\ 
	& \bs{\hat{x}}_{k+N|k} \in \mathcal{X}_\text{f}
	\end{alignat}
\end{subequations}
where $\mathbb{N}_{a:b} $ denotes the set of all integers between and including $a$ and $b$. The optimization minimizes the cost function with respect to all predicted inputs. Therefore, all inputs are combined in one input sequence vector
\begin{align}
\bs{U}_k = \begin{bmatrix}
\bs{\hat{u}}_{k|k}  \\ \bs{\hat{u}}_{k+1|k} \\ \vdots \\ \bs{\hat{u}}_{k+N-1|k} 
\end{bmatrix} \in \mathcal{U}^N \subset \mathbb{R}^{mN}
\end{align}
and $\bs{U}_k^*$ is the optimal solution for the input sequence. 
The power notations of a set means the $N$-times Cartesian products of the set, i.e.,  $\mathcal{U}^N = \mathcal{U} \times \mathcal{U} \times ...$ .
Furthermore, in \eqref{eq:opt} the terminal cost $\bs{Q}_\text{f}\in\mathbb{R}^{n\times n}$ is used to achieve stability (see Section \ref{ssec:stability}).  The recursive feasibility is a property of MPC, which ensures that the optimization problem is solvable at each time step. The recursive feasibility is here achieved by the terminal constraint $\mathcal{X}_\text{f}\subseteq \mathcal{X}$ (see Section \ref{ssec:rf}).

\subsubsection{Trajectory Tracking}
MPC can also be used to steer the state towards a given trajectory $\bs{x}_{k,\text{ref}}$ with a reference input $\bs{u}_{k,\text{ref}}$. Then, the optimal control problem penalizes the deviation from the reference trajectory, i.e.,
\begin{subequations}\label{eq:opt_tr}
	\begin{align}
	\bs{U}_k^* =&  \arg \min_{\bs{U}_k} J_\text{tt}( \bs{\hat{x}}_{k|k}, \bs{U}_k) \\
	\begin{split}
	=&  \arg \min_{\bs{U}_k} \sum_{i=0}^{N-1} \left[ \left(\bs{\hat{x}}_{k+i|k}-\bs{x}_{k+i,\text{ref}}\right)^\top \bs{Q} \left(\bs{\hat{x}}_{k+i|k}-\bs{x}_{k+i,\text{ref}}\right)+ \left(\bs{\hat{u}}_{k+i|k}-\bs{u}_{k+i,\text{ref}}\right)^\top \bs{R}\left(\bs{\hat{u}}_{k+i|k}-\bs{u}_{k+i,\text{ref}}\right)\right] \\
		&\hspace{1.35cm} + \left(\bs{\hat{x}}_{k+N|k}-\bs{x}_{k+N,\text{ref}}\right) ^\top \bs{Q}_\text{f}\left(\bs{\hat{x}}_{k+N|k}-\bs{x}_{k+N,\text{ref}}\right) 
	\end{split}\\
	 &\hspace{0.5cm} \text{s.t.}\quad   \bs{\hat{x}}_{k+i+1|k}=\bs{A}\bs{\hat{x}}_{k+i|k}+\bs{B}\bs{\hat{u}}_{k+i|k}   \qquad i \in \mathbb{N}_{0:N-1}  \\
	&\hspace{1.35cm}  \bs{\hat{u}}_{k+i|k} \in \mathcal{U} \hspace{3.8cm} i \in \mathbb{N}_{0:N-1} \\
	&\hspace{1.35cm}  \bs{\hat{x}}_{k+i|k} \in \mathcal{X} \hspace{3.8cm} i \in \mathbb{N}_{0:N-1}  \\ 
	&\hspace{1.35cm}  \bs{\hat{x}}_{k+N|k} \in \mathcal{X}_\text{f}
	\end{align}
\end{subequations}
The regulation problem \eqref{eq:opt} is a special case of the trajectory tracking in \eqref{eq:opt_tr} with $\bs{x}_{k,\text{ref}}=0$ and  $\bs{u}_{k,\text{ref}}=0$.
Ideally, the reference trajectory $\bs{x}_{k,\text{ref}},\bs{u}_{k,\text{ref}}$ should be a solution of \eqref{eq:sys}. Then the optimization is only concerned with the deviation of the reference. 
Otherwise, a mismatch between the reference trajectory and the solution of \eqref{eq:opt_tr} is inevitable. For example, if an arbitrary reference trajectory is chosen for the state but a zero input reference is used, i.e., $\bs{u}_{k,\text{ref}}=0$, the optimization must find a compromise between fulfilling the state or input reference since both cannot be satisfied simultaneously. Additionally, with a reference trajectory satisfying \eqref{eq:sys}, stability can be shown using the method from Section \ref{ssec:stability}.

\subsection{Model Predictive Controller}
At each time instance $k$, the current state $\bs{x}_k$ of the system \eqref{eq:sys} is measured. This measurement is used as the basis for the state prediction, i.e.,
\begin{align}
	\bs{\hat{x}}_{k|k} = \bs{x}_k
\end{align}
Based on $\bs{\hat{x}}_{k|k}$ the optimal control problem with the cost function $J( \bs{\hat{x}}_{k|k}, \bs{U}_k^*) $  from \eqref{eq:opt} or the cost function $J_\text{tt}( \bs{\hat{x}}_{k|k}, \bs{U}_k) $  from \eqref{eq:opt_tr} is applied to get the optimal input sequence $\bs{U}_k^*$. 
The control law for the current time step is then the first element of the optimal input sequence, i.e., 
\begin{align}
\bs{u}_k = \kappa(\bs{x}_k) = \bs{\hat{u}}_{k|k}^*
\end{align}
All remaining inputs are not used in the control law. 
This input $\bs{u}_k$ is applied to system \eqref{eq:sys}. 
At the next time step $k+1$  the procedure is repeated based on the measurement  of the subsequent state $\bs{x}_{k+1}$.

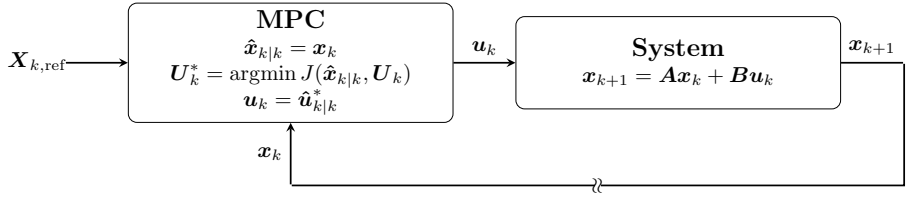
\begin{figure}[ht]
	\label{fig:mpc}
	\centering
	\tikzstyle{block} = [rectangle,
rounded corners, 
minimum width  = 3cm,
minimum height = 1.5cm,
draw = black]
\tikzstyle{arrow} = [thick,->,>=stealth]
\tikzstyle{empty}=[inner sep=0,outer sep=0]
\tikzstyle{dot}=[circle,inner sep = 0.05cm,draw =black,fill = black]
\centering
\resizebox{0.7\textwidth}{!}{
\begin{tikzpicture}[node distance = 1cm]
\node(mpc)[block]{\parbox{5cm}{\centering {\large \textbf{MPC}}\\ $\bs{\hat{x}}_{k|k}=\bs{x}_k$  \\ $\bs{U}_k^* =\arg\!\min J( \bs{\hat{x}}_{k|k}, \bs{U}_k)$ \\ $\bs{u}_k = \bs{\hat{u}}_{k|k}^*$ }};
\node(sys) [block, right = of mpc ]{\parbox{5cm}{\centering {\large\textbf{System}} \\$\bs{x}_{k+1} =\bs{A}\bs{x}_{k}+\bs{B}\bs{u}_{k}$}};
\node(h1)[empty,right = of sys]{};
\node(h2)[empty,below  = of  mpc]{};
\node(h3)[empty,left = of mpc]{$\bs{X}_{k,\text{ref}}$};

\draw[arrow](mpc)--node[anchor=south] {$\bs{u}_k$} (sys);
\draw[arrow](sys)--node[anchor=south] {$\bs{x}_{k+1}$} (h1)
				 |-node[pos=0.75,fill=white,rotate=90,inner sep=-1.25pt,outer sep=0,anchor=center]{$\approx$}(h2)
				 --node[anchor=east] {$\bs{x}_k$} (mpc);
\draw[arrow](h3)-- node[anchor=south] {}(mpc);
\end{tikzpicture}}
	\caption{MPC scheme}
\end{figure}

\subsection{Recursive Feasibility - Terminal Constraint}\label{ssec:rf}
Recursive feasibility is an important property for MPC because it guarantees if the optimization problems in \eqref{eq:opt} or \eqref{eq:opt_tr} have a solution, that the subsequent optimization are also feasible. Examples where recursive feasibility is not ensured are given in Section \ref{sec:simRF} or in \cite[Example 7.1]{GP17}.
In this tutorial, recursive feasibility is achieved by using terminal constraints. Recursively feasible MPC methods without terminal constraint are presented in \cite{GP17}.
If the terminal set is a control invariant set \cite{BBM17}, i.e.,
\begin{align}\label{eq:civs}
\bs{x} \in\mathcal{X}_\text{f} \subseteq \mathcal{X}  \implies \exists \bs{u}\in\mathcal{U} \quad\text{s.t.}\quad  \bs{A}\bs{x}+\bs{B}\bs{u} \in \mathcal{X}_\text{f},
\end{align}
then recursive feasibility is guaranteed. 
The reason for this is, that the solution of an optimization leads to an optimal state trajectory $ $  
\begin{align}
\bs{X}_k^* = \begin{bmatrix}
\bs{\hat{x}}_{k|k}^* \\ \bs{\hat{x}}_{k+1|k}^*\\ \vdots \\\bs{\hat{x}}_{k+N|k}^*
\end{bmatrix}  \in \mathcal{X}^{N+1} \subseteq \mathbb{R}^{n(N+1)},
\end{align}
where the last state is in the terminal set, i.e.,  $ \bs{\hat{x}}_{k+N|k}^* \in \mathcal{X}_\text{f} $. 
Due to the application of the first optimal input  $\bs{u}_k=\bs{\hat{u}}_{k|k}^*$ the initial state of the next optimization is the subsequent state of the prediction, i.e.,
\begin{align}
\bs{x}_{k+1} = \bs{\hat{x}}_{k+1|k}.
\end{align}
Therefore, a feasible (but not necessarily optimal) solution for the optimization at the next time step is a shifted version of the optimal solution of the first optimization, i.e.,
\begin{align}
\bs{X}_{k+1} =
\begin{bmatrix}
\bs{\hat{x}}_{k+1|k+1} \\
\bs{\hat{x}}_{k+2|k+1} \\
\vdots \\
\bs{\hat{x}}_{k+N|k+1} \\
\bs{\hat{x}}_{k+N+1|k+1} 
\end{bmatrix} =  
\begin{bmatrix}
\bs{\hat{x}}_{k+1|k}^* \\
\bs{\hat{x}}_{k+2|k}^* \\
\vdots \\
\bs{\hat{x}}_{k+N|k}^* \\
\bs{\tilde{x}}
\end{bmatrix}.
\end{align}

Since a shifted version of the previous solution is used for this consideration, a new state is necessary for the last predicted state $\bs{\tilde{x}}$. The state $\bs{\hat{x}}_{k+N|k}^*$ is in the terminal set $\mathcal{X}_\text{f}$. Due to \eqref{eq:civs}, an input exists such that $\bs{\tilde{x}}$ is feasible. Therefore the method is recursive feasible if \eqref{eq:civs} holds.

The largest possible control invariant set is called maximum invariant set and for more information the reader is referred to \cite{BBM17}.
The smallest possible control invariant set is the origin, i.e., $\mathcal{X}_\text{f} = \left\{\bs{0}\right\}$.

However, a large terminal set is preferable because a smaller terminal set also shrinks  the set of feasible initial states 
\begin{align}
\mathcal{X}_N = \left\{  \bs{\hat{x}}_{k|k} \in \mathcal{X} \;\middle|\;   \bs{\hat{x}}_{k+i|k} \in \mathcal{X}, \quad  \bs{\hat{u}}_{k+i|k} \in \mathcal{U}  \quad \forall i \in \mathbb{N}_{0:N-1},\qquad \bs{\hat{x}}_{k+N|k} \in \mathcal{X}_\text{f}  \right\} .
\end{align}
The set of feasible initial states $\mathcal{X}_N$ contains all initial states, for which the MPC method is feasible. 

\emph{Remark 1:} 
If the terminal set $\mathcal{X}_\text{f}$ is the maximum invariant set,  the set of feasible initial states is also the maximum invariant set, i.e., $\mathcal{X}_N=\mathcal{X}_\text{f}$. 

\emph{Remark 2:}
More practical  is the maximum stabilizing set of the origin \cite[Def. 11.13]{BBM17}. It contains all states, that can be steered to the origin. 
For this set it also holds, that the set of feasible initial states is also the maximum stabilization set i.e., $\mathcal{X}_N=\mathcal{X}_\text{f}$, but additionally the origin is always feasible.
The computation of the maximum stabilizing set is shown in Section \ref{ssec:compTermSet}.

\emph{Remark 3:} 
If the terminal set is the origin, i.e., $\mathcal{X}_\text{f}=\left\{\bs{0}\right\}$, the set of initial states $\mathcal{X}_N$ increases with an increasing horizon $N$.

\subsection{Stability - Terminal Cost}\label{ssec:stability}
For stability, it is important that the method is always solvable. Therefore, recursive feasibility from Section \ref{ssec:rf} is necessary for stability.  
Furthermore, stability of the optimal control problem in \eqref{eq:opt} can be shown, if the terminal cost $\bs{Q}_\text{f}$ is chosen appropriately. How to choose  $\bs{Q}_\text{f}$ is shown in the following. 

For this purpose, the cost function is used as Lypunov function, since the optimal cost is decreasing in the next step, i.e.,
\begin{align}
	J( \bs{\hat{x}}_{k|k}, \bs{U}_k^*) \geq 	J( \bs{\hat{x}}_{k+1|k+1}, \bs{U}^*_{k+1}) 
\end{align}
where $J( \bs{\hat{x}}_{k|k}, \bs{U}^*)$ is the cost function of the optimization in \eqref{eq:opt}, i.e.,
\begin{align}\label{eq:cost}
		J( \bs{\hat{x}}_{k|k}, \bs{U}^*_k) = \sum_{i=0}^{N-1} \left[\bs{\hat{x}}_{k+i|k}^\top \bs{Q} \bs{\hat{x}}_{k+i|k}+ \bs{\hat{u}}_{k+i|k}^\top \bs{R} \bs{\hat{u}}_{k+i|k} \right ] + \bs{\hat{x}}_{k+N|k} ^\top \bs{Q}_\text{f} \bs{\hat{x}}_{k+N|k}  
\end{align}
and $	J( \bs{\hat{x}}_{k+1|k+1},  \bs{U}^*_{k+1}) $ is the cost function in the next time instance where the controller is applied. The input sequence  $\bs{U}^*_{k+1}$ represents the optimal solution of the subsequent optimization but it is not known. However, the shifted version of the input sequence $\bs{U}_k^*$, i.e., 
\begin{align}\label{eq:uplus} 
 \bs{\tilde{U}}_{k+1} =\begin{bmatrix}
  \bs{\hat{u}}_{k+1|k}\\ \bs{\hat{u}}_{k+2|k}\\ \vdots \\ \bs{\hat{u}}_{k+N-1|k} \\ \bs{K}\bs{\hat{x}}_{k+N|k}
 \end{bmatrix}
\end{align}
with a stabilizing state feedback matrix $\bs{K}$ is a known and feasible solution but leads to a higher cost as the optimal solution, i.e. ${	J( \bs{\hat{x}}_{k+1|k+1}, \bs{U}^*_{k+1})<	J( \bs{\hat{x}}_{k+1|k+1}, \bs{\tilde{U}}_{k+1}) }$.  Nevertheless, the origin is stable if
\begin{align}\label{eq:de}
	J( \bs{\hat{x}}_{k+1|k+1}, \bs{\tilde{U}}_{k+1})  - J( \bs{\hat{x}}_{k|k}, \bs{U}^*_{k}) \leq 0 	
\end{align}
holds. With the definition of the cost function \eqref{eq:cost} and the subsequent input sequence vector \eqref{eq:uplus}  the difference in \eqref{eq:de} is 
\begin{align}
\begin{split}
J( \bs{\hat{x}}_{k+1|k+1}, \bs{\tilde{U}}_{k+1})  - J( \bs{\hat{x}}_{k|k}, \bs{U}^*_{k})= &	
-\bs{\hat{x}}_{k|k}^\top \bs{Q} \bs{\hat{x}}_{k|k} - \bs{\hat{u}}_{k|k}^\top \bs{R}\bs{\hat{u}}_{k|k} \\
&+\bs{\hat{x}}_{k+N|k}^\top \bs{Q} \bs{\hat{x}}_{k+N|k}+ \bs{\hat{u}}_{k+N|k}^\top \bs{R}\bs{\hat{u}}_{k+N|k}  \\
& - \bs{\hat{x}}_{k+N|k} ^\top \bs{Q}_\text{f} \bs{\hat{x}}_{k+N|k}\\
& + \bs{\hat{x}}_{k+N+1|k} ^\top \bs{Q}_\text{f} \bs{\hat{x}}_{k+N+1|k}
\end{split}
\end{align}
By assuming a LQR feedback law $\bs{u}_k = \bs{K}\bs{x}_k$  further simplifications are possible since the control law yields the closed loop system $\bs{x}_{k+1} = \left(\bs{A}+\bs{B}\bs{K}\right)\bs{x}_k$
\begin{subequations}
\begin{align}
\begin{split}
J( \bs{\hat{x}}_{k+1|k+1}, \bs{\tilde{U}}_{k+1})  - J( \bs{\hat{x}}_{k|k}, \bs{U}^*_{k})= &
-\bs{\hat{x}}_{k|k}^\top \left(\bs{Q} +\bs{K}^\top  \bs{R}\bs{K}\right)\bs{\hat{x}}_{k|k} \\
&+\bs{\hat{x}}_{k+N|k}^\top \left(\bs{Q} +\bs{K}^\top  \bs{R}\bs{K}\right)\bs{\hat{x}}_{k+N|k}  \\
& - \bs{\hat{x}}_{k+N|k} ^\top \bs{Q}_\text{f} \bs{\hat{x}}_{k+N|k}\\
& + \bs{\hat{x}}_{k+N|k} ^\top \left(\bs{A}+\bs{B}\bs{K}\right)^\top\bs{Q}_\text{f}\left(\bs{A}+\bs{B}\bs{K}\right)\bs{\hat{x}}_{k+N|k}
\end{split}\\
\begin{split}
= &
\underbrace{
-\bs{\hat{x}}_{k|k}^\top \left(\bs{Q} +\bs{K}^\top  \bs{R}\bs{K}\right)\bs{\hat{x}}_{k|k} }_{\leq 0} \\
&+\bs{\hat{x}}_{k+N|k}^\top \left(\bs{Q} +\bs{K}^\top  \bs{R}\bs{K}-\bs{Q}_\text{f}  + \left(\bs{A}+\bs{B}\bs{K}\right)^\top\bs{Q}_\text{f}\left(\bs{A}+\bs{B}\bs{K}\right) \right)\bs{\hat{x}}_{k+N|k}  
\end{split}
\end{align}
\end{subequations}
Therefore, the difference \eqref{eq:de} is negative or zero if
\begin{align}\label{eq:dare}
	\bs{Q} +\bs{K}^\top  \bs{R}\bs{K}-\bs{Q}_\text{f} 	 + \left(\bs{A}+\bs{B}\bs{K}\right)^\top\bs{Q}_\text{f}\left(\bs{A}+\bs{B}\bs{K}\right) = \bs{0}
\end{align}
holds. This is the discrete-time Riccati equation and the solution for $\bs{Q}_\text{f}$ leads to a terminal cost, which stabilizes the MPC controller. 

The feedback matrix $\bs{K}$ is determined with LQR control \cite{PLB15} and is given by
\begin{align}
	\bs{K} = \left(\bs{B}^\top \bs{Q}_\text{f}\bs{B}+\bs{R}\right)^{-1}\left(\bs{B}^\top\bs{Q}_\text{f}\bs{A}\right)
\end{align}
In MATLAB, the terminal cost matrix $\bs{Q}_\text{f}$ and the feedback matrix $\bs{K}$ can be computed with the command \texttt{[Qf,K,$\sim$] = idare(A,B,Q,R,[],[])}, an implicit solver of the discrete-time algebraic Riccati equation.

\section{Implementation}

The implementation in MATLAB of MPC is shown in the following. The optimal control problem in \eqref{eq:opt}  is modified to become a quadratic program, which can be solved with the MATLAB function \texttt{quadprog} or \texttt{mpcActiveSetSolver}.

\subsection{Lifted System Dynamics}
The recursive manner of the optimal control problem \eqref{eq:opt} can be solved with a nested function and a non-linear solver such as \texttt{fmincon}. But for a linear system it is preferable to use a quadratic solver, since the computation is faster. 
Therefore, the notation of a lifted system dynamics
\begin{align}\label{eq:liftedSys}
\bs{X}_k = \bs{\mathcal{A}} \bs{x}_k + \bs{\mathcal{B}} \bs{U}_k,
\end{align}
is used, where the whole state sequence can be determined with the aid of the input sequence $\bs{U}_k$ for  a given initial state $\bs{x}_k$. The state sequence and the input sequence are 
\begin{align}\label{eq:stateinputseq}
\bs{X}_k = \begin{bmatrix}
\bs{\hat{x}}_{k|k} \\ \bs{\hat{x}}_{k+1|k} \\ \vdots \\ \bs{\hat{x}}_{k+N|k}
\end{bmatrix}  \in \mathcal{X}^{N+1} \subseteq \mathbb{R}^{n(N+1)},
\qquad 
\bs{U}_k = \begin{bmatrix}
\bs{\hat{u}}_{k|k} \\ \bs{\hat{u}}_{k+1|k}\\ \vdots \\ \bs{\hat{u}}_{k+N-1|k}
\end{bmatrix}  \in \mathcal{U}^{N} \subset \mathbb{R}^{mN}
\end{align}
respectively. 
The lifted system matrix and the lifted input matrix are 
\begin{align}
\bs{\mathcal{A}}=
\begin{bmatrix}
\bs{I} \\ \bs{A} \\ \bs{A}^2 \\ \vdots \\ \bs{A}^N
\end{bmatrix} \in \mathbb{R}^{n(N+1)\times n},
\qquad
\bs{\mathcal{B}}=
\begin{bmatrix}
\bs{0} & \bs{0} & \hdots & \bs{0}\\ 
\bs{B} & \bs{0} & \hdots & \bs{0}\\ 
\bs{A}\bs{B} & \bs{B} &  \hdots & \bs{0}\\ 
\vdots &\vdots &&\vdots \\ 
\bs{A}^{N-1}\bs{B} &\bs{A}^{N-2}\bs{B}&\vdots& \bs{B}
\end{bmatrix} \in \mathbb{R}^{n(N+1)\times m(N+1)}
\end{align}
respectively. 

\subsection{Constraints}

The inputs must be within the input set $\bs{u}_{k}\in\mathcal{U}$ and the states must be within the state set $\bs{u}_{k}\in\mathcal{X}$ for all time steps. 
Both sets are convex and contain at least the origin. 
For the computation, the sets are defined as H-polyhedra \cite{BBM17}, i.e., the sets are defined  as inequality constraints. 
The state set is defined as 
\begin{align}\label{eq:xset}
\mathcal{X} = \left\{ \bs{x}\in \mathbb{R}^n  \;\middle|\; \bs{F} \bs{x} \leq \bs{f} \right\}
\end{align}
and the input set is defined as
\begin{align}\label{eq:uset}
\mathcal{U} = \left\{ \bs{u} \in \mathbb{R}^m\;\middle|\; \bs{G} \bs{u} \leq \bs{g} \right\}
\end{align}
where \bs{F} and \bs{G} are matrices and \bs{f} and \bs{g} are vectors. The $\leq$ is here used element-wise. 
Additionally, the terminal state set is also necessary for the MPC optimization. This set is given by
\begin{align}\label{eq:fset}
\mathcal{X}_\text{f} = \left\{ \bs{x}\in \mathbb{R}^n \;\middle|\; \bs{F}_\text{f} \bs{x} \leq \bs{f}_\text{f} \right\}.
\end{align}

\emph{Example:} If a box constraint for state
$\bs{x}_k = \begin{bmatrix}x_1 &x_2 \end{bmatrix}^\top$ 
is used, where only values $|x_{1,2}|\leq 10$ are possible, the set is 
\begin{align}
	\mathcal{X} = \left\{ \bs{x} \in \mathbb{R}^n\;\middle|\; 
	\begin{bmatrix}
	1 & 0 \\ -1 &0 \\ 0 & 1 \\ 0 & -1
	\end{bmatrix}
	 \bs{x} \leq \begin{bmatrix}
	 10\\ 10 \\ 10 \\ 10
	 \end{bmatrix} \right\}.
\end{align}

\subsubsection{Lifted Constraints}
For the numeric solver in MATLAB, a constraint for the input sequence $\bs{U}_k$ is needed. Therefore, the constraints \eqref{eq:xset}, \eqref{eq:uset}, and  \eqref{eq:fset} must be reformulated. 
First, the constraints must also be lifted to have a constraint for the state and input sequence \eqref{eq:stateinputseq}.
The state sequence is constraint with 
\begin{align}\label{eq:stateSecSet}
 \bs{\tilde{F}} \bs{X}_k \leq \bs{\tilde{f}} 
\end{align}
where the matrices are 
\begin{align}
\bs{\tilde{F}} =
\begin{bmatrix}
\bs{F}  & \bs{0} & \hdots & \bs{0}& \bs{0}  \\
\bs{0}     & \bs{F}  &  \hdots & \bs{0} & \bs{0} \\
\vdots & \vdots &&\vdots&\vdots \\
\bs{0}     & \bs{0} &  \hdots & \bs{F} & \bs{0}  \\
\bs{0}     & \bs{0} &  \hdots & \bs{0} & \bs{F}_f  
\end{bmatrix}  
\qquad 
\bs{\tilde{f}} =
\begin{bmatrix}\bs{f}\\\bs{f}\\\vdots\\\bs{f} \\\bs{f}_f	\end{bmatrix}
\end{align}
and input sequence is constraint with 
\begin{align}\label{eq:inputSecSet}
\bs{\tilde{G}} \bs{U}_k \leq \bs{\tilde{g}}
\end{align}
where the matrices are 
\begin{align}
	\bs{\tilde{G}} =
	\begin{bmatrix}
	\bs{G}  & \bs{0} & \hdots & \bs{0} \\
	\bs{0}     & \bs{G}  &  \hdots & \bs{0} \\
	\vdots & \vdots &&\vdots \\
	\bs{0}     & \bs{0} &  \hdots & \bs{G}  
	\end{bmatrix} 
	\qquad 
	\bs{\tilde{g}} =
	\begin{bmatrix}\bs{g}\\\bs{g}\\\vdots \\\bs{g}\end{bmatrix} .
\end{align}

\subsubsection{Admissible Input Set}

Since the optimization takes only the input sequence $\bs{U}_k$ into account, the state constraint \eqref{eq:stateSecSet} is transformed into the input space with the lifted system dynamics \eqref{eq:liftedSys}, i.e.,
\begin{subequations}
	\begin{align}
	\bs{\tilde{F}} \left(\bs{\mathcal{A}} \bs{x}_k + \bs{\mathcal{B}} \bs{U}_k\right) \leq \bs{\tilde{f}}  \\
	\bs{\tilde{F}}  \bs{\mathcal{B}} \bs{U}_k \leq \bs{\tilde{f}} -\bs{\tilde{F}}\bs{\mathcal{A}} \bs{x}_k  
	\end{align}
\end{subequations}

Therefore, the admissible input set is 
\begin{align}
\mathcal{U}_\text{ad}(\bs{x}_k) = \left\{ \bs{U}_k \;\middle|\; 
\begin{bmatrix}
\bs{\tilde{F}} \bs{\mathcal{B}} \\ \bs{\tilde{G}}
\end{bmatrix}
\bs{U}_k \leq 
\begin{bmatrix}
\bs{\tilde{f}} \\ \bs{\tilde{g}}\end{bmatrix} - 
\begin{bmatrix}
\bs{\tilde{F}} \bs{\mathcal{A}}
\\\bs{0}
\end{bmatrix} \bs{x}_k
\right\},
\end{align}
which depends on the initial state $\bs{x}_k$.

\subsubsection{Feasible Sets}
The set of feasible initial states can be expressed as
\begin{align}
\mathcal{X}_N = \left\{ \bs{x}_k \;\middle|\; \exists \bs{U}_k : 
\begin{bmatrix}
\bs{\tilde{F}} \bs{\mathcal{A}}
\\\bs{0}
\end{bmatrix} \bs{x}_k+
 \begin{bmatrix}
\bs{\tilde{F}} \bs{\mathcal{B}} \\ \bs{\tilde{G}}
\end{bmatrix}
\bs{U}_k
 \leq 
\begin{bmatrix}
\bs{\tilde{f}} \\ \bs{\tilde{g}}\end{bmatrix} 
\right\}.
\end{align}
The computation of it can be done with a Fourier-Motzkin projection, which can be efficiently computed with the MPT3 toolbox \cite{MPT3}.

\subsubsection{Computation of the Terminal Constraint}\label{ssec:compTermSet}
The terminal constraint $\mathcal{X}_\text{f}$ needs to be control invariant  \eqref{eq:civs}. The calculation of the terminal constraint requires some concepts from the set algebra, defined in \cite{BBM17}, i.e., the Minkovsky sum 
\begin{align}
	\mathcal{X}\oplus\mathcal{Y}=\left\{ \bs{x}+\bs{y}\;\middle|\;\bs{x}\in\mathcal{X},\bs{y}\in\mathcal{Y} \right\} ,
\end{align}
affine mappings of a set with a matrix, i.e., 
\begin{align}
	\bs{A}\circ\mathcal{X} = \left\{ \bs{y} \;\middle|\; \bs{y}=\bs{A}\bs{x}, \bs{x}\in\mathcal{X} \right\}  ,
\end{align} 
and the inverse affine mapping, i.e.,
\begin{align}
	\mathcal{X}\circ\bs{A} = \left\{ \bs{x} \;\middle|\;  \bs{A}\bs{x} \in \mathcal{X} \right\} . 
\end{align} 
The computation of this operations are implemented in the MPT3 toolbox~\cite{MPT3}.

The stabilizable set $\mathcal{K}_i$ of the origin is a control invariant set, which can be calculated with the following iterative algorithm \cite[Alg. 11.3]{BBM17}:
\begin{subequations}
\begin{align}
	\mathcal{K}_0 &= \left\{\bs{0}\right\}\\
	\mathcal{K}_{i+1} &= \left( \mathcal{K}_i \oplus \left(-\bs{B} \circ \mathcal{U}\right) \right) \circ \bs{A}  \cap \mathcal{X}.
\end{align}
\end{subequations}
The states in $\mathcal{K}_{i+1}$ represent all states, for which an input exist such that the subsequent state is in $\mathcal{K}_i$.
Since the origin is control invariant, it holds for all $i$ that $\mathcal{K}_{i}$ is control invariant and  $\mathcal{K}_{i}\subseteq\mathcal{K}_{i+1}$.
If the procedure converges, the resulting set is the maximal stabilizing set of the origin.

\subsection{Quadratic Program}
The implementation with the MATLAB built-in functions  \texttt{quadprog} or \texttt{mpcActiveSetSolver} are shown here. 
\subsubsection{Regulation Problem}
The optimization in \eqref{eq:opt} can be rewritten with the matrix notation \eqref{eq:stateinputseq}, which yields
\begin{subequations}\label{eq:opt2}
	\begin{alignat}{2}
	\bs{U}_k^* = & \arg \min_{\bs{U}_k} \bs{X}_k^\top\bs{\tilde{Q}}
	\bs{X}_k + \bs{U}_k^\top \bs{\tilde{R}} \bs{U}_k  \\
	\text{s.t.}\quad  & \bs{X}_k = \bs{\mathcal{A}} \bs{x}_k + \bs{\mathcal{B}} \bs{U}_k \\
	&\bs{U}_k\in\mathcal{U}_\text{ad}(\bs{x}_k) 
	\end{alignat}
\end{subequations}
with
\begin{align}
\bs{\tilde{Q}} =\begin{bmatrix}\bs{Q}&&&\\&\ddots&&\\&&\bs{Q}&\\&&&\bs{Q}_\text{f}\end{bmatrix}
 \in \mathbb{R}^{n(N+1)\times n(N+1)},
\qquad
\bs{\tilde{R}}=
\begin{bmatrix}\bs{R}&&\\&\ddots&\\&&\bs{R}\end{bmatrix}
\in \mathbb{R}^{mN\times mN}.
\end{align}
The optimization \eqref{eq:opt2} can be simplified by substitution of $\bs{X}_k$ with \eqref{eq:liftedSys}  yielding
\begin{subequations}\label{eq:opt3}
	\begin{alignat}{2}
	\bs{U}_k^* = & \arg \min_{\bs{U}_k}
	\bs{U}_k^\top \left(\bs{ \bs{\mathcal{B}}^\top\bs{\tilde{Q}}\bs{\mathcal{B}} +\tilde{R}}\right)  \bs{U}_k  + 2 \bs{x}_k^\top\bs{\mathcal{A}}^\top \bs{\mathcal{B}}\bs{U}_k + \bs{x}_k^\top\bs{\mathcal{A}}^\top \bs{\tilde{Q}}\bs{\mathcal{A}}\bs{x}_k
	 \\
	\text{s.t.}\quad  &\bs{U}_N\in\mathcal{U}_\text{ad}(\bs{x}_k) .
	\end{alignat}
\end{subequations}
The last part of the objective function is constant with respect to $\bs{U}_k$. Therefore, it can be neglected. The MATLAB built-in function solves the minimization
\begin{align}
	\arg\min_{\bs{x}} \bs{x}^\top \bs{H} \bs{x} + 2 \bs{f}^\top \bs{x}.
\end{align} 
Therefore the variables for \texttt{quadprog} or \texttt{mpcActiveSetSolver} are
\begin{subequations}
\begin{align}
	\bs{H} & = \bs{ \bs{\mathcal{B}}^\top\bs{\tilde{Q}}\bs{\mathcal{B}} +\tilde{R}} \\
	\bs{f} & = \bs{\mathcal{B}}^\top \bs{\mathcal{A}} \bs{x}_k. 
\end{align}
\end{subequations}

\subsubsection{Trajectory Tracking}
The optimization with reference trajectory \eqref{eq:opt_tr} can be reformulated with the lifted system dynamics \eqref{eq:liftedSys} and the stacked notation of the reference trajectories 
\begin{align}
	\bs{X}_{k,\text{ref}} =\begin{bmatrix}
	\bs{x}_{k,\text{ref}} \\
	\bs{x}_{k+1,\text{ref}} \\
	\vdots\\
	\bs{x}_{k+N,\text{ref}}
	\end{bmatrix}
	\in \mathcal{X}^{N+1} \subseteq \mathbb{R}^{n(N+1)},
	\qquad 
	\bs{U}_{k,\text{ref}} = \begin{bmatrix}
	\bs{\hat{u}}_{k,\text{ref}} \\ \bs{\hat{u}}_{k+1,\text{ref}}\\ \vdots \\ \bs{\hat{u}}_{k+N-1,\text{ref}}
	\end{bmatrix}  \in \mathcal{U}^{N} \subset \mathbb{R}^{mN}
\end{align}
 to 
\begin{subequations}
	\begin{alignat}{2}
	\bs{U}_k^* = & \arg \min_{\bs{U}_k} \left(\bs{X}_k-\bs{X}_{k,\text{ref}}\right)^\top\bs{\tilde{Q}}
	\left(\bs{X}_k-\bs{X}_{k,\text{ref}}\right) + \left(\bs{U}_k-\bs{U}_{k,\text{ref}}\right)^\top \bs{\tilde{R}} \left(\bs{U}_k-\bs{U}_{k,\text{ref}}\right)  \\
	\text{s.t.}\quad  & \bs{X}_k = \bs{\mathcal{A}} \bs{x}_k + \bs{\mathcal{B}} \bs{U}_k \\
	&\bs{U}_k\in\mathcal{U}_\text{ad}(\bs{x}_k) 
	\end{alignat}
\end{subequations}

The cost function can be simplified with the substitution of $\bs{X}_k$ with \eqref{eq:liftedSys} such that it is solvable for the MATLAB solver, i.e.,
\begin{subequations}
	\begin{align}
	J_\text{tt}(\bs{x}_k,\bs{U}_k)&= \left(\bs{X}_k-\bs{X}_{k,\text{ref}}\right)^\top\bs{\tilde{Q}}
	\left(\bs{X}_k-\bs{X}_{k,\text{ref}}\right) + \left(\bs{U}_k-\bs{U}_{k,\text{ref}}\right)^\top \bs{\tilde{R}} \left(\bs{U}_k-\bs{U}_{k,\text{ref}}\right) \\
	&=\bs{U}_k^\top  \left(\bs{\mathcal{B}}^\top\bs{\tilde{Q}}\bs{\mathcal{B}} + \bs{\tilde{R}} \right) \bs{U}_k + 2 \left(\bs{x}_k^\top\bs{\mathcal{A}}^\top\bs{\tilde{Q}}\bs{\mathcal{B}} -\bs{X}_{k,\text{ref}}^\top\bs{\tilde{Q}}\bs{\mathcal{B}} -\bs{U}_{k,\text{ref}}^\top \bs{\tilde{R}}\right)\bs{U}_k  \\
	&+ \underbrace{\bs{x}_k^\top\bs{\mathcal{A}}^\top\bs{\tilde{Q}}\bs{\mathcal{A}} \bs{x}_k -
		2\bs{x}_k^\top\bs{\mathcal{A}}^\top\bs{\tilde{Q}}\bs{X}_{k,\text{ref}}+
		\bs{X}_{k,\text{ref}}^\top\bs{\tilde{Q}}\bs{X}_\text{ref} +\bs{U}_{k,\text{ref}}^\top \bs{\tilde{R}}\bs{U}_{k,\text{ref}}}_{\text{const.}}
	\end{align}
\end{subequations}

The constant term does not affect the optimization, since it is constant with respect to the decision variable~$\bs{U}_k$.
Therefore, the constant part can be neglected and the optimization for the MATLAB solver is
\begin{subequations}
	\begin{alignat}{2}
	\bs{U}_k^* = & \arg \min_{\bs{U}_k}\bs{U}_k^\top
	\underbrace{\left(\bs{\mathcal{B}}^\top\bs{\tilde{Q}}\bs{\mathcal{B}} + \bs{\tilde{R}} \right)}_{\bs{H}}\bs{U}_k + 2\underbrace{\left(\bs{x}_k^\top\bs{\mathcal{A}}^\top\bs{\tilde{Q}}\bs{\mathcal{B}} -\bs{X}_{k,\text{ref}}^\top\bs{\tilde{Q}}\bs{\mathcal{B}} -\bs{U}_{k,\text{ref}}^\top \bs{\tilde{R}}\right)}_{\bs{f}^\top}\bs{U}_k   \\
	\text{s.t.}\quad  &\bs{U}_k\in\mathcal{U}_\text{ad}(\bs{x}_k) 
	\end{alignat}
\end{subequations}
In each evaluation, the optimization depends on the current measurement of the state $\bs{x}_k$.

\section{Example}
In this section, an example of MPC is shown. The system is defined as
\begin{align}
\bs{A}=\begin{bmatrix}1&T\\0&1\end{bmatrix}, \qquad \bs{B} = \begin{bmatrix}0\\T\end{bmatrix}
\end{align}
where $T$ is the discretization time step and is chosen as $T=0.05$. 
The states are two dimensional, i.e., $\bs{x}_k=\begin{bmatrix}x_1&x_2\end{bmatrix}^\top \in \mathbb{R}^2$ and the input is one dimensional, i.e., $\bs{u}_k=u\in \mathbb{R}$.
The states and inputs are constrained with box constraints, i.e.,
\begin{align}
\mathcal{X} = \left\{ \bs{x} \;\middle|\; 
\begin{bmatrix}
1 & 0 \\ -1 &0 \\ 0 & 1 \\ 0 & -1
\end{bmatrix}
\bs{x} \leq \begin{bmatrix}
10\\ 10 \\ 10 \\ 10
\end{bmatrix} \right\},
 \quad
\mathcal{U} = \left\{ u\;\middle|\; 
\begin{bmatrix}
1 \\ -1 
\end{bmatrix}
u \leq \begin{bmatrix}
20\\ 20 
\end{bmatrix} \right\}.
\end{align}
The cost matrices are chosen as 
\begin{align}
	\bs{Q}= \begin{bmatrix}	1&0\\0&1\end{bmatrix},\qquad R = 1.
\end{align} 
The terminal cost $\bs{Q}_\text{f}$ is the solution of the discrete-time Riccati equation \eqref{eq:dare}. For solving the equation, the \texttt{idare} command is used, resulting in 
\begin{align}
\bs{Q}_\text{f}=\begin{bmatrix}	35.7&20.9\\20.9&36.2	\end{bmatrix}.
\end{align} 

In the following, three simulations are shown. The first is a simple example of MPC, where the state is steered to the origin. 
In the second example, the loss of feasibility is shown. This problem arises because the state reaches a point in the state space from which no solution of the optimization exists. 
Therefore, the last simulation shows the same situation but with a terminal constraint to avoid the loss of feasibility. 

\subsection{Regulation Problem}
In the first simulation, the reference for the control method is the origin, i.e., a regulation problem is demonstrated. The MPC horizon is chosen as $N=10$. 
\begin{figure}[ht!]
	\centering
	\input{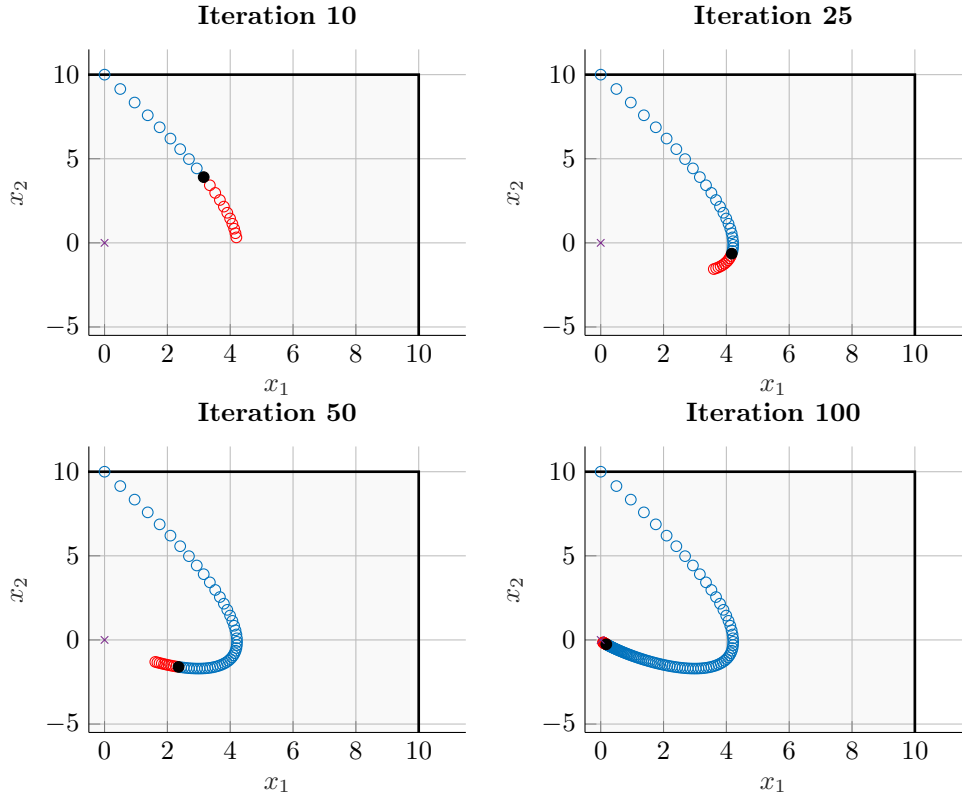}
	\caption{MPC regulation problem; blue: previous states, black: current state, red: MPC prediction}
	\label{fig:ex1}
\end{figure}

Figure~\ref{fig:ex1} shows the simulation results with the initial state $\bs{x}_0=\begin{bmatrix}0&10\end{bmatrix}^\top$. The current state $\bs{x}_k$ is shown as a black dot. The blue circles are the previous states and the red circle are the predicted states based on the optimal input sequence $\bs{U}_k^*$.
The state constraint $\mathcal{X}$ is denoted as a gray box.

In each iteration of MPC, the states are predicted (red) based on the current state (black). Then, the first input of the optimal input sequence is applied to the system. 

\subsection{Loss of Recursive Feasibility}\label{sec:simLossRF}
In this section, the need of the recursive feasibility is demonstrated. In the simulation, a horizon of $N=5$ and the initial state $\bs{x}_0=\begin{bmatrix}7.3&10\end{bmatrix}^\top$ is used. However, this simulation does not use the terminal set $\mathcal{X}_\text{f}$. 
\begin{figure}[ht!]
	\centering
%
%
\definecolor{mycolor1}{rgb}{0.50196,0.50196,0.50196}%
\definecolor{mycolor2}{rgb}{0.00000,0.44700,0.74100}%
\definecolor{mycolor3}{rgb}{0.49400,0.18400,0.55600}%
\begin{tikzpicture}

\begin{axis}[%
width=1.952in,
height=1.528in,
at={(0.758in,0.206in)},
scale only axis,
xmin=-0.5,
xmax=11.5,
xlabel style={font=\color{white!15!black}},
xlabel={$x_1$},
ymin=-5.5,
ymax=11.5,
ylabel style={font=\color{white!15!black}},
ylabel={$x_2$},
axis background/.style={fill=white},
title style={font=\bfseries},
title={Iteration 4},
axis x line*=bottom,
axis y line*=left,
xmajorgrids,
ymajorgrids
]

\addplot[area legend, line width=1.0pt, draw=black, fill=mycolor1, fill opacity=0.05, forget plot]
table[row sep=crcr] {%
x	y\\
10	-10\\
10	10\\
-10	10\\
-10	-10\\
}--cycle;
\addplot [color=mycolor2, only marks, mark=o, mark options={solid, mycolor2}, forget plot]
  table[row sep=crcr]{%
7.24	10\\
7.74	9\\
8.19	8\\
};
\addplot [color=red, only marks, mark=o, mark options={solid, red}, forget plot]
  table[row sep=crcr]{%
8.94	6\\
9.24	5.05982732082783\\
9.49299136604139	4.18553642952278\\
9.70226818751753	3.37375497493583\\
9.87095593626432	2.62124071990706\\
};
\addplot [color=black, only marks, mark=*, mark options={solid, fill=black, black}, forget plot]
  table[row sep=crcr]{%
8.59	7\\
};
\addplot [color=mycolor3, only marks, mark=x, mark options={solid, mycolor3}, forget plot]
  table[row sep=crcr]{%
0	0\\
};
\end{axis}

\begin{axis}[%
width=1.952in,
height=1.528in,
at={(3.327in,0.206in)},
scale only axis,
xmin=-0.5,
xmax=11.5,
xlabel style={font=\color{white!15!black}},
xlabel={$x_1$},
ymin=-5.5,
ymax=11.5,
ylabel style={font=\color{white!15!black}},
ylabel={$x_2$},
axis background/.style={fill=white},
title style={font=\bfseries},
title={Iteration 5},
axis x line*=bottom,
axis y line*=left,
xmajorgrids,
ymajorgrids
]

\addplot[area legend, line width=1.0pt, draw=black, fill=mycolor1, fill opacity=0.05, forget plot]
table[row sep=crcr] {%
x	y\\
10	-10\\
10	10\\
-10	10\\
-10	-10\\
}--cycle;
\addplot [color=mycolor2, only marks, mark=o, mark options={solid, mycolor2}, forget plot]
  table[row sep=crcr]{%
7.24	10\\
7.74	9\\
8.19	8\\
8.59	7\\
};
\addplot [color=red, only marks, mark=o, mark options={solid, red}, forget plot]
  table[row sep=crcr]{%
9.24	5.05405510543854\\
9.49270275527193	4.17573285194874\\
9.70148939786936	3.36165084605247\\
9.86957194017199	2.60856119656027\\
10	1.91334748285587\\
};
\addplot [color=black, only marks, mark=*, mark options={solid, fill=black, black}, forget plot]
  table[row sep=crcr]{%
8.94	6\\
};
\addplot [color=mycolor3, only marks, mark=x, mark options={solid, mycolor3}, forget plot]
  table[row sep=crcr]{%
0	0\\
};
\end{axis}
\end{tikzpicture}%
	\caption{Loss of feasibility; blue: previous states, black: current state, red: MPC prediction
	\label{fig:ex2}}
\end{figure}
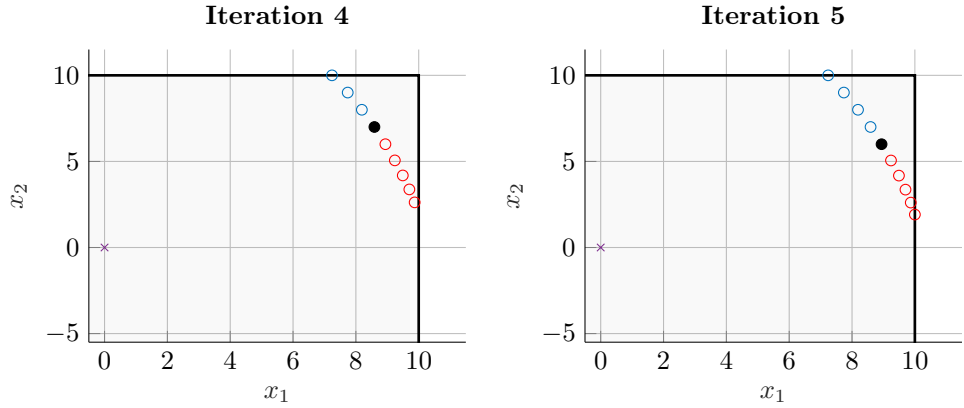

In the first 5 iterations of the MPC method are feasible as shown in Figure~\ref{fig:ex2}. However, the last predicted state $\bs{\hat{x}}_{k+N|k} = \bs{\hat{x}}_{9|4}$ is already on the edge of the state constraint $\mathcal{X}$. In the next time step, there exists no solution for the optimization, since there is no input such that the state does not violate the constraint. Therefore, the solver raises an error and the simulation stops. 

This can be avoided with the use of the terminal constraint, which is shown in the next section. 

\subsection{Recursive Feasibility through Terminal Constraint}\label{sec:simRF}
The malfunction from the previous simulation is avoidable if a suitable terminal constraint $\mathcal{X}_\text{f}$ is used (see. Section \ref{ssec:rf}).
The maximum stabilizing set, computed with the MPT3 toolbox, is
\begin{align}\label{eq:terminaSet}
	\mathcal{X}_\text{f} = \left\{ \bs{x} \;\middle|\; 
	\begin{bmatrix}
	-0.999&	-0.050 \\
	0.999&	0.050\\
	0.995&	0.100\\
	0.989&	0.148\\
	0.981&	0.196\\
	0.970&	0.243\\
	0.958&	0.287\\
	0.944&	0.330\\
	0.928&	0.371\\
	0.912&	0.410\\
	0.894&	0.447
	\end{bmatrix}
	\bs{x} \leq \begin{bmatrix}
	10.0\\
	10.0\\
	10.0\\
	10.0\\
	10.1\\
	10.2\\
	10.3\\
	10.4\\
	10.6\\
	10.8\\
	11.0
	\end{bmatrix} \right\} \cap \mathcal{X}.
\end{align}

\begin{figure}[ht!]
	\centering
	\input{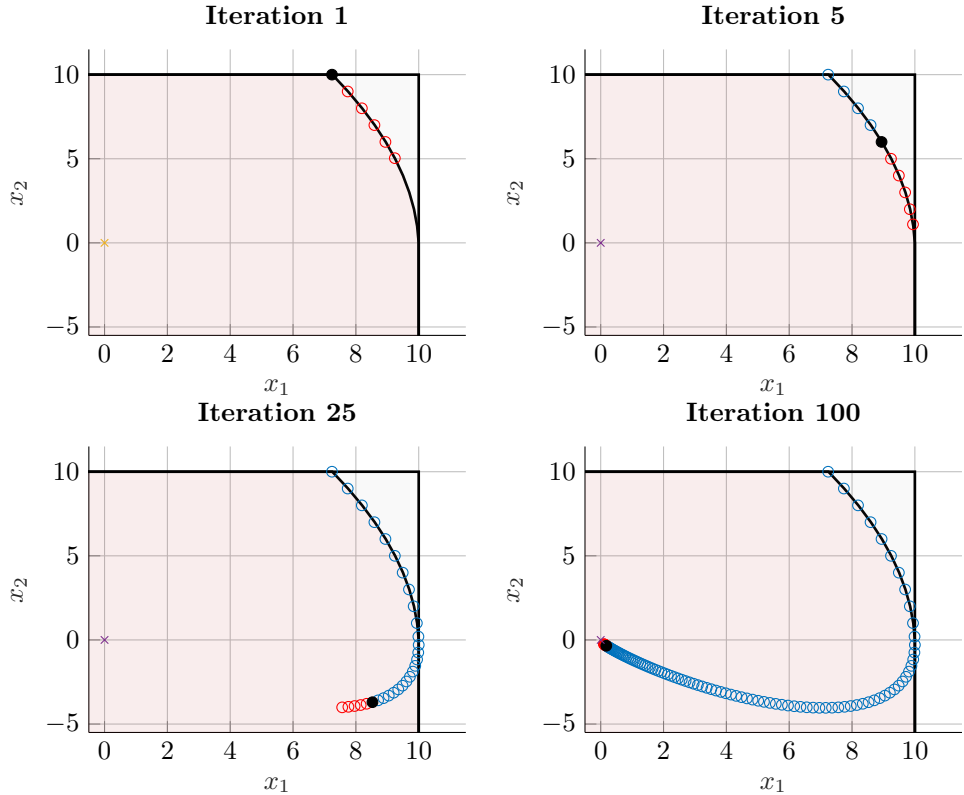}
	\caption{Recursive feasibility; blue: previous states, black: current state, red: MPC prediction, reddish area: terminal set $\mathcal{X}_\text{f}$}
	\label{fig:ex3}
\end{figure}

\begin{figure}[ht!]
	\centering
%
%
\definecolor{mycolor1}{rgb}{0.00000,0.44700,0.74100}%
\begin{tikzpicture}

\begin{axis}[%
width=4.521in,
height=1.643in,
at={(0.758in,0.38in)},
scale only axis,
xmin=-1,
xmax=100,
xlabel style={font=\color{white!15!black}},
xlabel={Iterations},
ymin=-21,
ymax=5,
ylabel style={font=\color{white!15!black}},
ylabel={$u$},
axis background/.style={fill=white}
]
\addplot[const plot, color=mycolor1, line width=1.0pt, forget plot] table[row sep=crcr] {%
0	-20\\
1	-20\\
2	-20\\
3	-20\\
4	-20\\
5	-20\\
6	-20\\
7	-20\\
8	-20\\
9	-15.9999999999997\\
10	-9.90806239600818\\
11	-9.07196028684224\\
12	-8.28350115300344\\
13	-7.5406202215154\\
14	-6.84131489302544\\
15	-6.18364437831197\\
16	-5.56572921696976\\
17	-4.98575068919965\\
18	-4.44195013097881\\
19	-3.93262816226648\\
20	-3.45614383730732\\
21	-3.0109137255294\\
22	-2.59541093099473\\
23	-2.20816405784766\\
24	-1.84775612871769\\
25	-1.51282346256898\\
26	-1.20205451804738\\
27	-0.914188707956115\\
28	-0.648015190093789\\
29	-0.402371639310139\\
30	-0.17614300527766\\
31	0.0317397398620217\\
32	0.222300860138826\\
33	0.396521117147148\\
34	0.555339030942828\\
35	0.699652137466701\\
36	0.830318239772471\\
37	0.948156650577445\\
38	1.05394942388117\\
39	1.14844257361039\\
40	1.23234727744899\\
41	1.30634106420006\\
42	1.37106898320349\\
43	1.42714475449814\\
44	1.47515189857245\\
45	1.51564484469211\\
46	1.54915001692854\\
47	1.57616689713818\\
48	1.5971690642598\\
49	1.61260520940633\\
50	1.62290012632913\\
51	1.62845567692644\\
52	1.62965173155487\\
53	1.62684708398316\\
54	1.6203803409012\\
55	1.61057078596586\\
56	1.59771921842722\\
57	1.58210876643644\\
58	1.56400567518846\\
59	1.54366007010025\\
60	1.52130669526874\\
61	1.49716562749106\\
62	1.47144296616511\\
63	1.44433149941976\\
64	1.41601134685179\\
65	1.38665057927156\\
66	1.35640581588098\\
67	1.3254227993263\\
68	1.29383694908453\\
69	1.26177389365594\\
70	1.22934998204694\\
71	1.19667277503661\\
72	1.16384151672817\\
73	1.13094758689174\\
74	1.09807493460915\\
75	1.06530049373367\\
76	1.03269458067878\\
77	1.0003212750494\\
78	0.968238783627897\\
79	0.936499788224187\\
80	0.905151777895834\\
81	0.87423736603941\\
82	0.843794592849056\\
83	0.813857213632043\\
84	0.784454973464367\\
85	0.755613868662065\\
86	0.727356395536\\
87	0.69970178688955\\
88	0.672666236709817\\
89	0.646263113493904\\
90	0.620503162642341\\
91	0.595394698342115\\
92	0.570943785351818\\
93	0.547154411091457\\
94	0.524028648429204\\
95	0.501566809547157\\
96	0.479767591257812\\
97	0.458628212132586\\
98	0.438144541793362\\
99	0.418311222707663\\
};
\end{axis}
\end{tikzpicture}%
	\caption{Applied input $u$}
	\label{fig:ex4}
\end{figure}
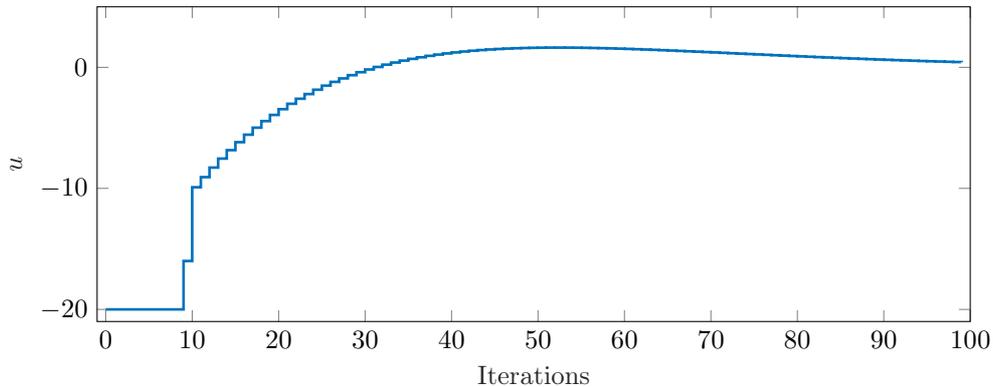

The same simulation from Figure \ref{fig:ex2} but with the terminal set from \eqref{eq:terminaSet} is demonstrated in Figure \ref{fig:ex3}. The last predicted state is always within the terminal set $\mathcal{X}_\text{f}$ and therefore, a subsequent state can always be found. 
The terminal set $\mathcal{X}_\text{f}$ is highlighted by the reddish area. Since it is the maximum stabilizing set, it coincides with the set of feasible initial states $\mathcal{X}_N$. The initial state $\bs{x}_0=\begin{bmatrix}7.3&10\end{bmatrix}^\top$  is on the edge of $\mathcal{X}_N$. Therefore, a slight change in the initial state can lead to leaving the set of feasible initial states and thus to the loss of feasibility.
The input $u$ applied in this simulation is shown in Figure \ref{fig:ex4}. In the first iterations, the input has the value $-20$ because the input constraint is active. After a few steps, the constraint is no longer active, and the optimization finds a sequence of inputs that fulfill the control task.
\section{Conclusion}
This tutorial gives an overview of linear MPC. However, in real applications, the control method must be able to handle disturbances. Disturbances affect stability and recursive feasibility. Therefore, there are several approaches that deal with disturbances. Robust MPC (RMPC) avoids violating constraints even in the presence of the worst disturbances. There are several RMPC approaches, a common one is tube-based RMPC \cite{ML01}. Another method is Stochastic MPC (SMPC). It utilizes the stochastic properties of the disturbances. A common SMPC approach is that a small probability of constraint violation is allowed. An overview of different SMPC methods is given in \cite{M16}.

\bibliographystyle{unsrt}
\bibliography{jobname} 
\end{document}